\documentclass[%
 reprint,
 amsmath,amssymb,
 aps,
]{revtex4-2}
\usepackage{graphicx}
\usepackage{dcolumn}
\usepackage{bm}
\usepackage{hyperref}
\begin{document}
\preprint{APS/123-QED}
\title{Gating of two-dimensional electron systems in InGaAs/InAlAs heterostructures:\\the role of the intrinsic InAlAs deep donor defects}
%
\author{Michael Prager}
\author{Michaela Trottmann}
\author{Jaydean Schmidt}
\author{Lucia Ebnet}
\author{Dieter Schuh}
\author{Dominique Bougeard}%
 \email{dominique.bougeard@ur.de}
\affiliation{%
 Institut f{\"u}r Experimentelle und Angewandte Physik, Universit{\"a}t Regensburg, D-93040 Regensburg, Germany
}%
\date{\today}
%
\begin{abstract}
We present an analysis of gated InGaAs/InAlAs heterostructures, a device platform to realize spin-orbitronic functionalities in semiconductors. The phenomenological model deduced from our magnetotransport experiments allows us to correlate the gate response of the studied two-dimensional electron systems to heterostructure design parameters, in particular the indium concentration. We explain the occurrence of metastable electrostatic configurations showing reduced capacitive coupling and provide gate operation strategies to reach classical field effect control in such heterostructures. Our study highlights the role of the intrinsic InAlAs deep donor defects, as they govern the dynamics of the electrostatic response to gate voltage variations through charge trapping and unintentional tunneling. 
\end{abstract}
%
\maketitle
%
%
\section{\label{sec:intro}Introduction}
Quantum wells (QW) based on InAlAs potential barriers have developed into a diverse platform to leverage spin-orbit interaction for solid-state applications. These QW heterostructures allow an engineering and control of the Rashba-type spin-orbit interaction \cite{grundler2000, nitta1997, choi2008, kohda2012, holmes2008, nitta2009, simmonds2009, kunihashi2009}, opening perspectives \cite{manchon2015} for the realization of all-electrical spin-transistors \cite{chuang2015} or topological superconductivity including the observation of Majorana zero modes \cite{shabani2016, suominen2017, lee2019, fornieri2019, frolov2020}.\\
Gating of two-dimensional carrier systems represents a key functionality of these applications and has been frequently used in this context \cite{nitta1997, choi2008, kohda2012, holmes2008, nitta2009, Capotondi.2005, Chen.2015, Desrat.2004, Hatke.2017, Richter.2000, Shabani.2014, Shabani.2014b}. Interestingly, reports on the microscopic mechanisms of the capacitive coupling between the gates and the two-dimensional systems are scarce, while at the same time similarly designed heterostructures seem to deliver diverse gate responses \cite{Capotondi.2005, Chen.2015, Desrat.2004, Hatke.2017, Richter.2000, Shabani.2014, Shabani.2014b}, posing the question of the origin of these discrepancies.\\ 
Here, we present an extensive experimental study of the gate response of top-gated two-dimensional electron systems (2DES) confined in InAlAs-embedded InGaAs QWs. From our experiments, we deduce a phenomenological microscopic model which allows to explain and predict the observed features of the gate response. In particular, we elucidate the occurrence of a limited voltage range allowing classical field effect control of the 2DES, while a broader range leads to metastable gate response, with limited to vanishing capacitive coupling of the gate to the 2DES. We also demonstrate gate operation strategies to bring an InAlAs-based heterostructure from a metastable situation into the classical field effect range.\\
Our model points out the crucial role of intrinsic InAlAs defect states for the electrostatics of gated heterostructures, hence highlighting their importance in the design process of InAlAs-containing devices for spin-orbitronic applications. 
\section{\label{sec:exp}Experimental}
Our samples were grown by molecular beam epitaxy (MBE) on a semi-insulating GaAs (100) substrate. The layout of the heterostructures is sketched in Fig.~\ref{fig:figure1}(a). The oxide desorption was performed at $620^{\circ}$C, followed by 100 nm GaAs and an Al$_\textrm{0.5}$Ga$_{\textrm{0.5}}$As/GaAs ten period superlattice as lattice-matched smoothing layers. Due to the native lattice mismatch of 7.2\% between GaAs and InAs, a In$_\textrm{x}$Al$_{\textrm{1-x}}$As step-graded buffer was grown at $335^{\circ}$C, where x $= 0.1$ to $0.87$ (sample A) in 19 steps resp. to $0.82$ (sample B) in 17 steps followed by a single step back to x $= 0.81$ (sample A) resp. x $= 0.75$ (sample B). Then, the active layers were grown at $440^{\circ}$C, comprising the In$_\textrm{x}$Ga$_{\textrm{1-x}}$As QW embedded in 120 nm In$_\textrm{x}$Al$_{\textrm{1-x}}$As below and 130 nm In$_\textrm{x}$Al$_{\textrm{1-x}}$As above. In order to prevent oxidation damage, the structure was completed with a 2 nm In$_\textrm{x}$Ga$_{\textrm{1-x}}$As barrier. The indium content in the active layers is constant at 81\% for sample A and 75\% for sample B.\\
The samples were processed with standard wet etching techniques in a clean room environment to define a Hall bar geometry with a width of 20 $\mu$m. After etching, Ohmic contacts were realized by deposition of 260 nm AuGe ($88\%/12\%$) and 66 nm Ni, followed by forming gas assisted annealing. As a gate dielectric, we deposited 100 nm (sample A) resp. 50 nm (sample B) Al$_2$O$_3$ via thermal atomic layer deposition at $300^{\circ}$C, followed by a 10/100 nm Ti/Au top gate (TG).\\
The measurements, unless otherwise stated, were carried out at a temperature of T $= 4.2$ K in a $^4$He dewar using standard low frequency lockin techniques at 17 Hz with an excitation current of 50 nA. Sample A yields a maximum electron mobility of $\mathrm{\mu} = 506000$ cm$^2$/Vs at a density of n $= 4.9\times 10^{11}$ cm$^{-2}$ with a cooldown TG voltage of V$_{\textrm{TG}} = 0$ V, Sample B gives $\mu = 258000$ cm$^2$/Vs at a density of n $= 7.5\times 10^{11}$ cm$^{-2}$. In time-resolved measurements of the electron density as a function of the gate voltage, the density was determined by applying a constant magnetic field B~$= 500$ mT while sweeping the gate voltage at sweep rates 1-5 mV/s. For all applied gate voltages, the Hall voltage at zero magnetic field yields $0$ V. Hence, the linear dependence between the Hall voltage and B gives the density. 
\section{\label{sec:results}Results}
\subsection{\label{ssec:gateresponse}Gate response}
\begin{figure}[!h]
\includegraphics{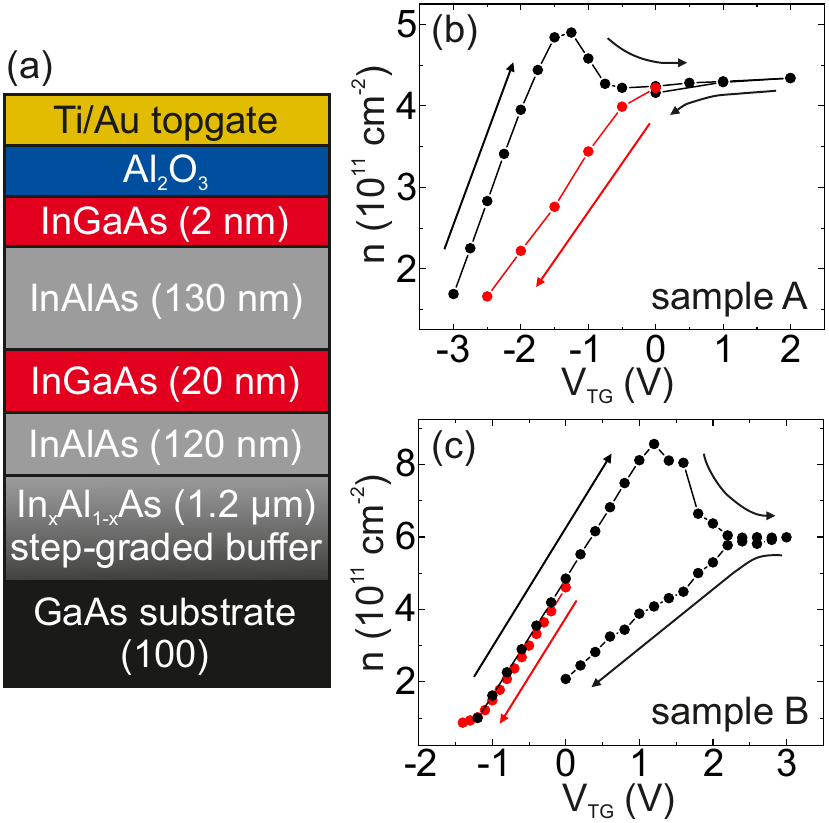}
\caption{\label{fig:figure1} Gate response of two representative samples A and B. (a) Heterostructure layout. The indium contents are 81\% for sample A and 75\% for sample B. (b) \& (c) Electron density n as a function of V$_{\textrm{TG}}$ for sample A (b) and sample B (c). After the cooldown to $4.2$ K at V$_{\textrm{TG}} = 0$ V, the system was depleted (red) below the MIT, followed by a TG upsweep and a subsequent downsweep back to V$_{\textrm{TG}} = 0$ V.}
\end{figure}
Fig.~\ref{fig:figure1}(b) displays the gate response, i.e. the electron density n of the 2DES as a function of the TG voltage V$_{\textrm{TG}}$, in an In$_\textrm{x}$Ga$_{\textrm{1-x}}$As QW embedded in In$_\textrm{x}$Al$_{\textrm{1-x}}$As with x $= 81\%$ (sample A). The applied gate voltage cycle was the following: the sample was cooled from room temperature (RT) to 4.2 K with V$_{\textrm{TG}} = 0$ V applied, yielding n $= 4.2\times 10^{11}$ cm$^{-2}$. The 2DES was then depleted by applying V$_{\textrm{TG}} < 0$V (red trace in Fig.~\ref{fig:figure1}(b)) until reaching the metal-insulator-transition (MIT). The latter occurred below n $= 1.8\times 10^{11}$ cm$^{-2}$ (last point of the red trace in Fig.~\ref{fig:figure1}(b)). After going to $-4$~V, V$_{\textrm{TG}}$ was increased: accumulation was then observed from V$_{\textrm{TG}} = -3$ V onwards (black trace in Fig.~\ref{fig:figure1}(b)). A clear hysteresis in the gate response appears, as the black trace differs from the red trace. Both branches exhibit a linear dependence, indicating a classical field effect response to the gate action. However, the black trace has a steeper slope than the red trace, indicating a stronger capacitive coupling of the TG to the 2DES in the QW. Further increasing V$_{\textrm{TG}}$ towards $0$ V, the linear part of the black trace ends at a peak density n$_{\textrm{peak}} = 4.9\times 10^{11}$ cm$^{-2}$. Then, although V$_{\textrm{TG}}$ is still increased, meaning electrons are accumulated in the QW, the measured density quickly decreases, forming a characteristic peak. While increasing V$_{\textrm{TG}}$ even further, this density drop saturates at n$_{\textrm{sat}} = 4.3\times 10^{11}$ cm$^{-2}$ and does not significantly change up to V$_{\textrm{TG}} = +2$V. Decreasing V$_{\textrm{TG}}$ again towards V$_{\textrm{TG}} = +1$ V results in no reaction of the 2DES density, indicating a loss of the capacitive coupling to the TG between $+2$ and $+1$ V. The 2DES reacts again below V$_{\textrm{TG}} = +1$ V: the electron density decreases (black trace in Fig.~\ref{fig:figure1}(b)). Its value at V$_{\textrm{TG}} = 0$ V precisely coincides with the density observed directly after the cooldown (merging of the black and the red trace in Fig.~\ref{fig:figure1}(b)). Finally, decreasing V$_{\textrm{TG}}$ into negative voltages exactly leads to the red trace in Fig.~\ref{fig:figure1}(b), closing the hysteresis loop. We observe this hysteresis loop to remain stable when repeating the described gate voltage cycle.\\
Regarding the efficiency of the gate control of the 2DES in view of device applications, two major observations hence emerge from Fig.~\ref{fig:figure1}(b): the sample allows a significant linear control of the 2DES density with the gate voltage, but this is true in two distinct gate voltage intervals with two different capacitive couplings (red and black trace). Furthermore, the maximal electron density that can be reached within these linear regimes is lower for the branch with the smaller capacitive coupling (red trace). The gate controllability can thus be expected to strongly depend on the electrostatic situation before the cooldown as well as on the gate sweep history of the sample. While this is rarely commented on in the literature, we find these observations to be experimentally in line with the state of the art: when compared, quite different gating properties are reported for similarly designed heterostructures \cite{Capotondi.2005, Chen.2015, Desrat.2004, Hatke.2017, Richter.2000, Shabani.2014, Shabani.2014b}, seemingly making a correlation between the gating behavior and the heterostructure layout elusive. Given that most of the previous reports used heterostructures with 75\% indium and to further illustrate this point, we show the gate response of sample B in Fig.~\ref{fig:figure1}(c): it is a 2DES in a device design which is identical to sample A (81\% indium), apart from the indium content in the semiconductor heterostructure, which has been reduced to 75\% here. While this sample B qualitatively yields the same hysteretic elements as sample A, three striking differences appear: First, when cooling down from RT at V$_{\textrm{TG}} = 0$ V, the 2DES is now initialized in the steeper linear gate control regime, while in sample A the regime with the stronger capacitive coupling (black trace in Fig.~\ref{fig:figure1}(b)) cannot be accessed when simply cooling the device down with V$_{\textrm{TG}} = 0$ V applied. Second, driving the 2DES in sample B beyond depletion (red trace in Fig.~\ref{fig:figure1}(c)) and then accumulating electrons again (black trace), leaves the 2DES unchanged, as the red and the black trace are identical. In sample A they were not. Finally, and third, the peak electron density $n_{\textrm{peak}} = 8.5\times 10^{11}$ cm$^{-2}$ as well as the saturation density n$_{\textrm{sat}} = 6.0\times 10^{11}$ cm$^{-2}$ are significantly higher in sample B compared to sample A.\\
In the following, we will develop a systematic understanding of the observed gating behavior in InGaAs/InAlAs heterostructures. Given that the 2DES properties of both samples strongly depend on the gate sweep history, we will start by discussing the reaction to a modification of the RT electrostatics of the samples.
\subsection{\label{ssec:biascd}Biased cooldown}
To modify the electrostatic environment at RT, we applied certain non-zero V$_{\textrm{TG}}$ at RT and then proceeded with the cooldown under the chosen non-zero bias, a procedure termed as biased cooldown (BCd) in the following.
\begin{figure}[!h]
\includegraphics{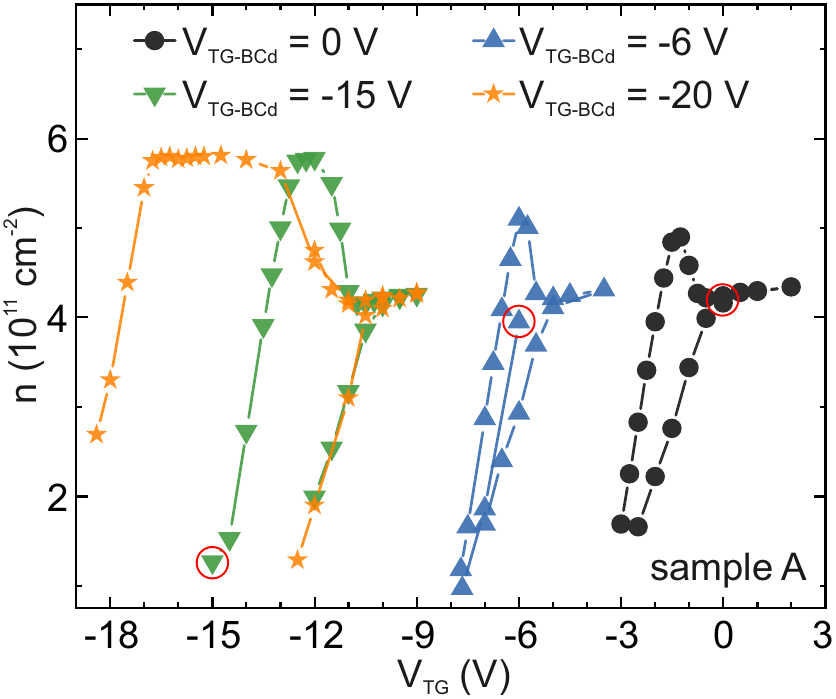}
\caption{\label{fig:figure2} Electron density n of sample A as a function of V$_{\textrm{TG}}$ for four representative biased cooldown voltages V$_{\textrm{TG-BCd}}$. The red circle indicates the first measurement point after cooldown at each V$_{\textrm{TG-BCd}}$.}
\end{figure}
Fig.~\ref{fig:figure2} displays the gate response of the 2DES in sample A after four different biased cooldown runs for which the gate voltage V$_{\textrm{TG-BCd}}$ applied at RT is decreased from $0$ V to $-20$ V. The starting point, which corresponds to the chosen V$_{\textrm{TG-BCd}}$ value, is circled in red for each run. The black curve with V$_{\textrm{TG-BCd}} = 0$ V is identical to Fig.~\ref{fig:figure1}(b) and thus starts in the linear regime with the smaller slope. Applying a negative RT bias V$_{\textrm{TG-BCd}} = -6$ V (blue trace) results in two important changes: First, after initializing the gate sweep loop at V$_{\textrm{TG-BCd}} = -6$ V, the depletion of the 2DES occurs linearly, but now with an intermediate slope. Second, the maximal electron density is increased compared to the black trace (V$_{\textrm{TG-BCd}} = 0$ V). Reducing V$_{\textrm{TG-BCd}}$ further, to $-15$ V (green trace), confirms this trend: now, the first measurement point is located in the linear regime with the steepest slope and the maximal electron density is further increased. Additionally, the peak, which concludes the steeper slope regime, is widened. Going even further, to V$_{\textrm{TG-BCd}} = -20$ V, the MIT already occurs at RT. When increasing V$_{\textrm{TG}}$ until electron accumulation (at $-18.4$ V), the system is initialized in the steeper linear branch, like for the green trace. While the maximal electron density does not further increase between the green and the orange trace, the peak, which concludes the steeper slope regime, is widened into a plateau. 
From our biased cooldown studies, which are reproducible across several samples and here shown exemplarily for sample A, we draw two conclusions: On a practical level, we here demonstrate from our very general experimental experience across many samples from different heterostructures, that proper biased cooling allows to initialize the gated heterostructure into a well-defined state, for example into the steepest linear slope. Strikingly, the gate response of the green trace in Fig.~\ref{fig:figure2} very much resembles the one of sample B (75\% indium and V$_{\textrm{TG-BCd}} = 0$ V), shown in Fig.~\ref{fig:figure1}(c). Thus, the fact that biased cooling of an 81\% indium sample allows to make the gated 2DES behave similarly to a zero-biased 75\% indium sample, suggests that the gate response of a sample is closely linked to the intrinsic microscopic electrostatics of the heterostructure. Hence, in the following, we will discuss and analyze the different parts of the gating cycles presented in Fig.~\ref{fig:figure1} and \ref{fig:figure2}.
\subsection{\label{ssec:ctm1}Charge transfer model for the gate response}
\begin{figure}[!h]
\includegraphics{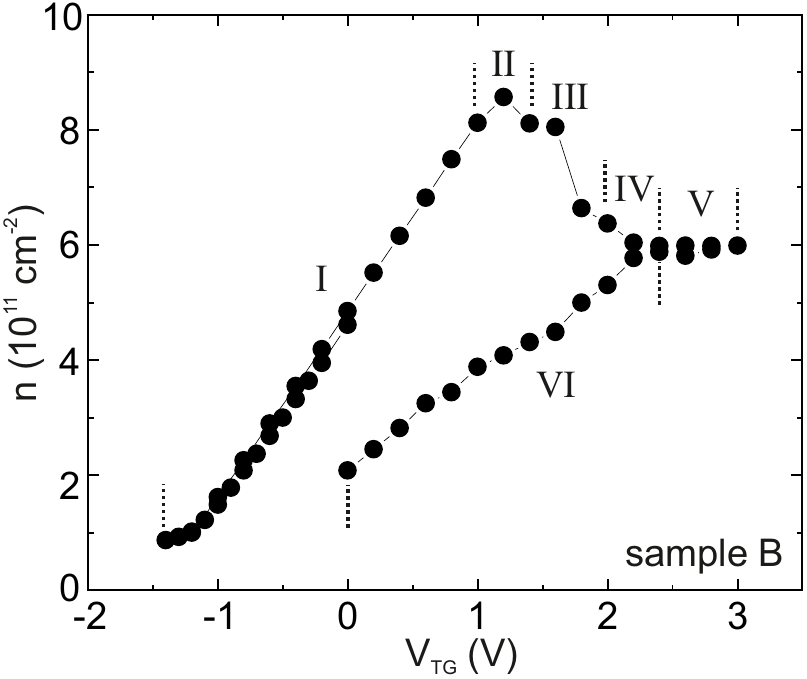}
\caption{\label{fig:figure3} Electron density n of sample B as a function of V$_{\textrm{TG}}$ at V$_{\textrm{TG-BCd}} = 0$ V. The different electrostatic regimes, which are discussed in the text, are indicated with roman numerals.}
\end{figure}
We divide the gate response hysteresis pattern into characteristic electrostatic regimes, numbered with roman numerals as presented in Fig.~\ref{fig:figure3}. These regimes represent reproducible elements of the hysteresis and do not necessarily coincide with the chronology of the V$_{\textrm{TG}}$ variation after cooling down. The following discussion and the presented charge transfer model (CTM) are supported by the experimental results from many samples, with e.g. different indium contents and different QW depths beyond $50$ nm below the heterostructure surface. Exemplarily we will use sample B to illustrate our discussion in this section.
\paragraph*{Regime I}
We define the steeper linear segment, which suggests capacitive coupling described by the classical field effect, as regime I. Experimentally, this regime is characterized by two major observations: First, up- and downsweeps of V$_{\textrm{TG}}$ coincide along this branch, and only along this branch: This will not be true in the following regimes. Second, when sweeping V$_{\textrm{TG}}$ to the next measurement value, the electron density instantly reaches its equilibrium value. This instant reaction is again specific to regime I. As we will see later, in other regimes the experimental determination of the electron density will require settling times for the electron density to reach its equilibrium value after each gate action.
\paragraph*{Regime II}
\begin{figure}[!htb]
\includegraphics{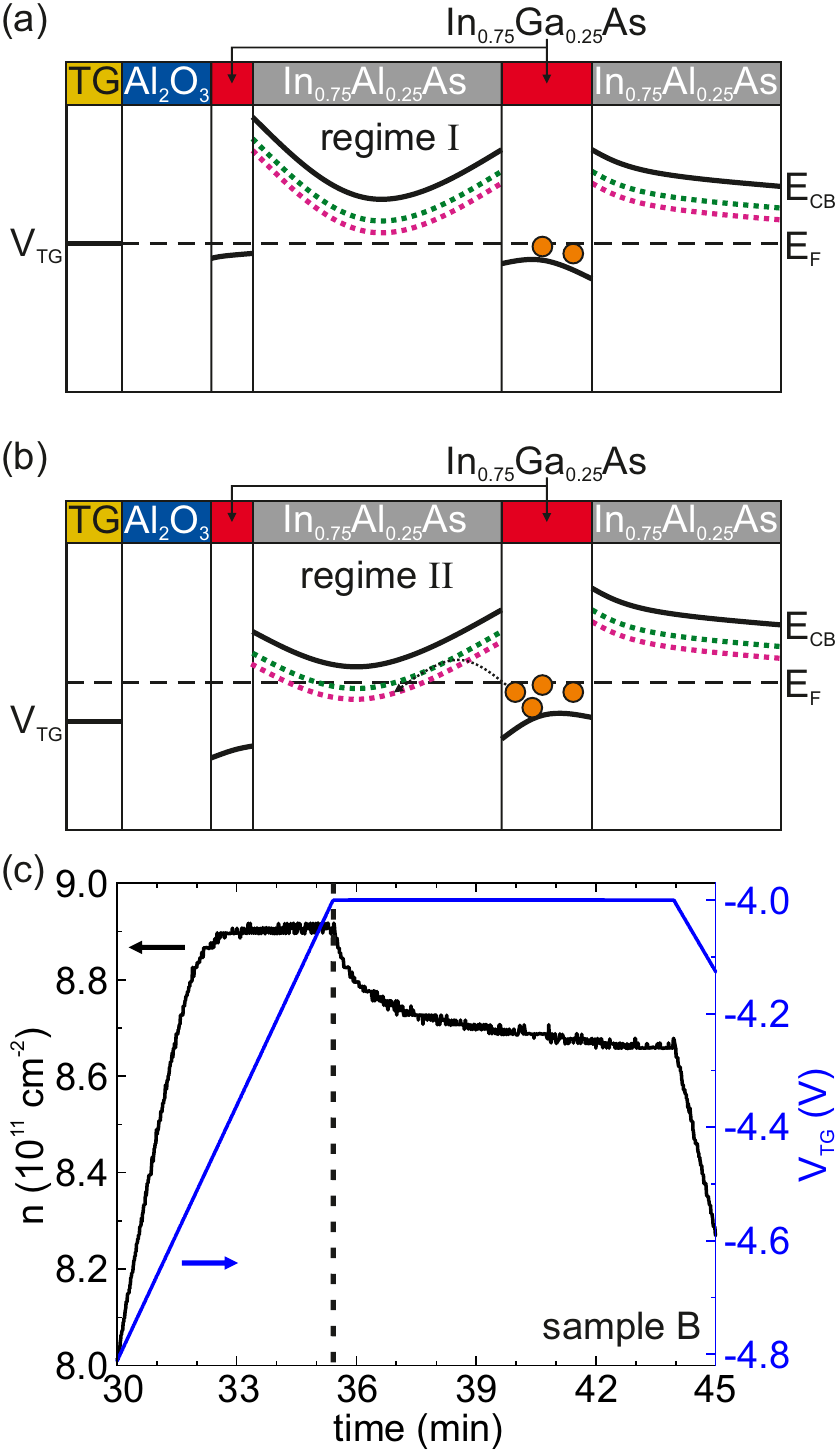}
\caption{\label{fig:figure4} (a) \& (b) CB edge sketch of sample B for V$_{\textrm{TG}} = 0$ V (a) and V$_{\textrm{TG}} > 0$ V (b), representative of regime I (a) and regime II (b). The color dotted lines indicate the energy of the DDLs within the InAs, lying at $0.12$ eV (green) and $0.17$ eV (pink) below the InAlAs CB edge. The orange dots represent an electron population. With increasing V$_{\textrm{TG}}$, electron tunneling from the QW into the InAlAs DDLs sets in. (c) Electron density n (black) of sample B and V$_{\textrm{TG}}$ sweep (blue) at V$_{\textrm{TG-BCd}} = -6$ V as a function of measurement time. The sample was initialized in regime I and evolves to regime II at minute 32.}
\end{figure}
Regime I has an upper limit, which appears either as a sharp or more elongated peak (even transforming into a plateau for V$_{\textrm{TG-BCd}} = -20$ V in Fig.~\ref{fig:figure2}), defined as regime II. Here, although V$_{\textrm{TG}}$ continues to be increased, the density saturates. This could either be a result of a loss of capacitive coupling between the gate and the 2DES or of a loss of electrons out of the QW. We exclude the loss of capacitive coupling, since the electron density continues to immediately react to a V$_{\textrm{TG}}$ reduction for all voltage values in regime II. While a loss of electrons out of the QW seems counterintuitive at first sight, it becomes more plausible when looking at the band edge sketches of the gated heterostructure in Fig.~\ref{fig:figure4}(a) and (b).
These schematics are inspired by self-consistent Schroedinger-Poisson calculations, for which we used the parameters from \cite{vurgaftman2001} and \cite{Capotondi.2004}.
Note, that during the MBE growth of InAlAs, the formation of intrinsic crystal defect sites in the form of arsenic antisites is inevitable. These arsenic antisites induce two deep donor level (DDL) states, which lie $0.12$ eV and $0.17$ eV respectively below the conduction band (CB) edge of InAlAs \cite{Capotondi.2004}. These quite effectively act as donor states for an InGaAs or InAs QW embedded in InAlAs, creating 2DESs with significant electron densities in these QWs through intrinsic doping, even in nominally undoped heterostructures. Fig.~\ref{fig:figure4}(a) depicts the CB edge (black lines) for sample B as well as the evolution of the two energy levels of the DDLs along the growth direction of the heterostructure (dotted colored lines). The electron population is schematically represented by orange dots. Due to the ionization of the DDLs, the evolution of the InAlAs CB edge along the growth direction is curved and, in particular, forms a trough-shape between the heterostructure surface and the InGaAs QW. In Fig.~\ref{fig:figure4}(a) V$_{\textrm{TG}} = 0$ V is applied to sample B, representative of the system initialized in regime I, in a stable configuration, as observed in the experiment. However, when applying positive V$_{\textrm{TG}}$, the band edge bends downwards in the part left of the QW in the sketch. From a certain V$_{\textrm{TG}} > 0$ V onwards (the start of regime II) this eventually results in the situation sketched in Fig.~\ref{fig:figure4}(b).
Compared to Fig.~\ref{fig:figure4}(a), the sketch illustrates that electrons in the QW now face an increasingly more transparent triangular-shaped potential barrier between the QW and the DDL states in the InAlAs. Simultaneously, the DDL energy is shifted to be equal or even below the lowest subband of the QW. Hence, an increased tunneling probability from the 2DES into the DDLs is expected. These electrons, which are then trapped in the DDLs in the InAlAs, do not contribute to the transport anymore and thus appear as lost 2DES electrons in our experiment.
The dynamics shown in Fig.~\ref{fig:figure4}(b) are experimentally verified by the measurement depicted in Fig.~\ref{fig:figure4}(c). Sample B (V$_{\textrm{TG-BCd}} = -6$ V) is first set to regime I at V$_{\textrm{TG}} = -4.8$ V, from which we sweep V$_{\textrm{TG}}$ into regime II up to $-4.0$ V while measuring the electron density of the 2DES over time. Regime I is represented by the instant reaction of the electron density to the classical field effect, i.e. increasing linearly with increasing V$_{\textrm{TG}}$. At V$_{\textrm{TG}} = -4.6$ V, the density increase then flattens, which we denote as the beginning of regime II. Further increasing V$_{\textrm{TG}}$ leads to a smooth transition into a saturation of the density (although V$_{\textrm{TG}}$ is increased). As soon as the V$_{\textrm{TG}}$ sweep is stopped within regime II, here exemplarily at $-4.0$ V, the density immediately decreases over a duration of several minutes, illustrating the loss of the 2DES electrons. Additionally Fig.~\ref{fig:figure4}(c) shows, that the capacitive coupling of the TG to the 2DES in the QW is not lost. At min $44$, V$_{\textrm{TG}}$ is decreased again, which results in an immediate decrease of the density, driven by the classical field effect. This experiment also shows, that a settling time on the scale of several minutes is required for the density to reach its equilibrium value after a specific V$_{\textrm{TG}}$ was set. Only in regime I, the measured density does not require this settling time. Note, that each density point of the measurements presented in Fig.~\ref{fig:figure1},~\ref{fig:figure2} and~\ref{fig:figure3}, hence includes an appropriate settling time on the scale of several minutes.
\paragraph*{Regime III}
\begin{figure}[!htb]
\includegraphics{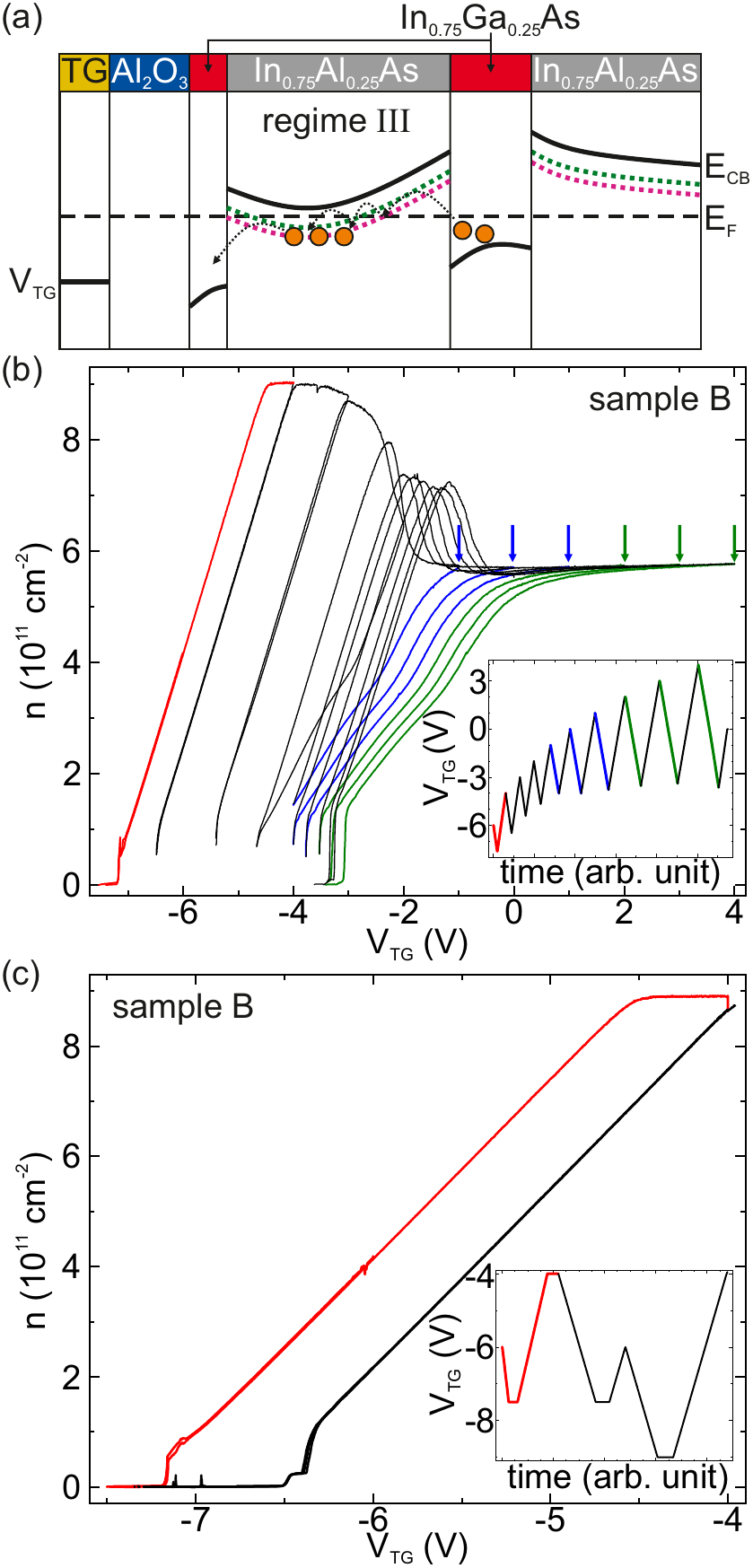}
\caption{\label{fig:figure5}
(a) CB edge sketch of sample B for increasing V$_{\textrm{TG}}$, representative of regime III. By increasing V$_{\textrm{TG}}$, electrons are transferred through the InAlAs via multi-step hopping tunneling. (b) Electron density n of sample B as a function of V$_{\textrm{TG}}$ at V$_{\textrm{TG-BCd}} = -6$ V. The red line indicates the starting point of the series. The details of the gate response series are discussed in the text. The inset depicts the V$_{\textrm{TG}}$ sweeps for the considered series. (c) Electron density n of sample B as a function of V$_{\textrm{TG}}$ at V$_{\textrm{TG-BCd}} = -6$ V for a 2DES depletion initiated respectively from regime I (red) and regime II (black). The inset depicts the corresponding V$_{\textrm{TG}}$ sweeps.
}
\end{figure}
By applying higher V$_{\textrm{TG}}$ at the regime II peak structure, the band tilting becomes even stronger. Hence, the triangular-shaped potential barrier between the QW and the InAlAs becomes even more transparent, effectively further increasing the tunneling rate of electrons from the QW to DDL sites within the InAlAs. From our experimental observations, we conclude that the electron tunneling rate becomes significantly larger than the increase of the electron density in the QW induced by the increasing V$_{\textrm{TG}}$ here, thus resulting in a clearly visible loss of electrons. In our model, this steep decrease of the 2DES density characterizes regime III.\\
Calculating the distance d between two InAlAs defect sites of the same type \footnote{Defect density N$_{\textrm{D}} \approx 3\times 10^{16}$ cm$^{-3}$, evenly divided between the two types. \cite{Brounkov.1995, Capotondi.2004}} yields d~$\approx 32$ nm. The effective Bohr radius of an electron bound to a doping site is calculated to a$_0^*\approx 18$ nm \footnote{$\mathrm{a}_0^* = \frac{\epsilon_{sc}}{m^*}\cdot\frac{4\pi\epsilon_0\hbar^2}{e^2}$, $\epsilon_{sc} = \epsilon_{In_{0.75}Al_{0.25}As} = 13.7$, $m^* = 0.041\cdot m_0$. \cite{trottmann2020}}. Comparing these two length scales d and a$_0^*$ suggests a field-assisted charge transfer through the InAlAs via multi-step hopping. This multi-step process manifests itself within the experiment, particularly in a comparatively long settling time of the electron density after each adjustment of V$_{\textrm{TG}}$. At the same time, when increasing V$_{\textrm{TG}}$, the apex of the trough is shifted towards the interface between the semiconductor (sc) and the dielectric. Hence, the increase of V$_{\textrm{TG}}$ will facilitate hopping from the trough apex towards this interface, as sketched in Fig.~\ref{fig:figure5}(a). In our understanding, when electrons arrive at this interface, they get trapped. Indeed, we have good experimental indications for this picture, as for example shown from the gate response in Fig.~\ref{fig:figure5}(b): Within a V$_{\textrm{TG}}$ variation in regime I (red trace), no hysteresis occurs, thus the downsweep to V$_{\textrm{TG}} = -7.5$ V and the following upsweep branch are identical. A reversion of V$_{\textrm{TG}}$ after leaving regime I (regimes II and onwards, black curve), at the contrary, results in the opening of a hysteresis, i.e. the branch of the gate voltage downsweep differs from the previous upsweep. The opening of this hysteresis reflects a loss of carriers in the 2DES, which we interpret as a transfer into the interface, where trapping occurs. Our experiments also show, that from regime II onwards this trapping at the interface is quite stable. Indeed, once an electron is trapped inside the interface, we observe that it cannot be brought back by a large negative V$_{\textrm{TG}}$. This is depicted in Fig.~\ref{fig:figure5}(c):  When regime II is reached (in this case in an upsweep to V$_{\textrm{TG}} = -4$ V) and the hysteresis occurs, neither a subsequent downsweep to V$_{\textrm{TG}} = -7.5$ V nor to V$_{\textrm{TG}} = -9$ V allows to reset the system back to the initial branch from regime I (red trace). In our experiments, only when heating up the sample to RT, the system undergoes the full reset.
\paragraph*{Role of the semiconductor-dielectric interface}
We believe that the observed stable trapping is most plausible at the interface and results from its complex morphology between the interface of the semiconductor heterostructure and the dielectric. First, since this interface marks the sharp transition between the single crystalline semiconductor heterostructure and the amorphous ALD-deposited Al$_2$O$_3$ (dielectric), it will host a significant concentration of point defects which provide an energetically broad distribution of states \cite{Choi.2013, Hoshii.2012, Lin.2011, Lin.2012, Robertson.2009, Taoka.2013, Wang.2011}, from shallow defects (which induce instabilities due to trap and release) to deep defects (which contribute to metastable trapping). Second, it additionally has been shaped by several environmental influences and sample processing steps - e.g. post-MBE growth oxidation of the InGaAs cap in air as well as a (possibly partial) self-cleaning of the oxidized surface at the beginning of the ALD process \cite{Ahn.2013, Hinkle.2009, Huang.2005, McIntyre.2009, Milojevic.2008, Park.2017, Shahrjerdi.2007, Timm.2010} - creating a spatially inhomogeneous, hummocky potential landscape at this interface.
\paragraph*{Regime IV}
\begin{figure}[!htb]
\includegraphics{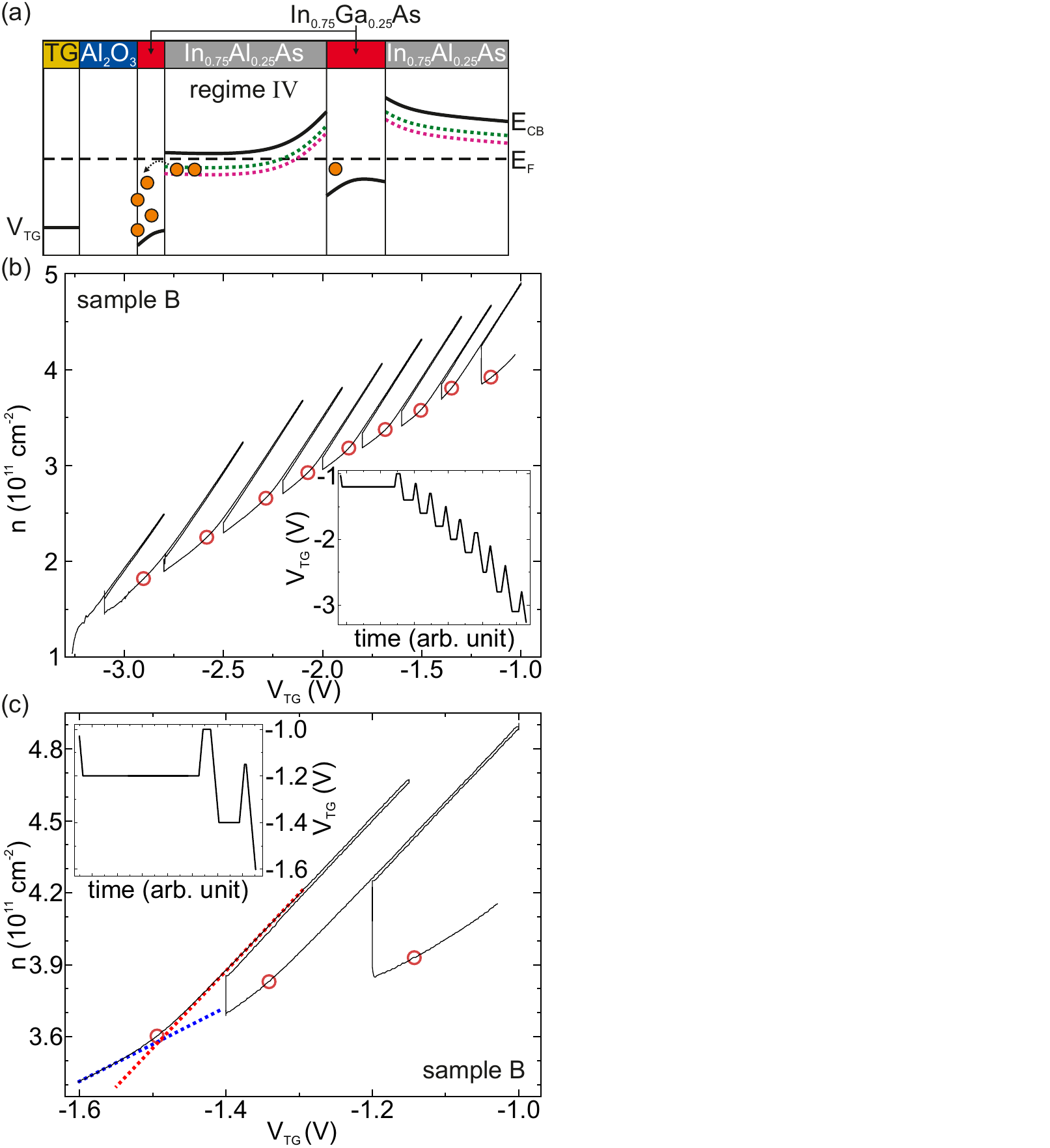}
\caption{\label{fig:figure6} (a) CB edge sketch of sample B for increasing (large) V$_{\textrm{TG}}$, representative of regime IV. The interface between the semiconductor and the dielectric gets populated. (b) Electron density n of sample B as a function of V$_{\textrm{TG}}$ at V$_{\textrm{TG-BCd}} = -6$ V within regime VI. The inset shows the applied V$_{\textrm{TG}}$ sweep routine which is discussed in the text. The red circles indicate the transition point of two slopes in each TG downsweep branch. (c) Zoom into (b): the red dotted line indicates the steeper slope value, the blue dotted line the smaller slope and the red circles the transition point between these two slope values.}
\end{figure}
In the experiments, the density drop in regime III ends with a smooth transition to a density saturation. The electron density then stays constant when further increasing V$_{\textrm{TG}}$. This regime is defined as regime IV. Its appearance indicates that when the interface hosts a sufficient threshold population of trapped negative charges, a partial screening of the electric field from the TG occurs. Indeed, this partial screening should induce a flattening of the band edge profile of the InAlAs compared to the unscreened situation, as depicted in Fig.~\ref{fig:figure6}(a). As a consequence, the triangular barrier between the QW and the InAlAs becomes less transparent again, effectively decreasing the tunneling rate from the QW to the InAlAs defect sites. The process which we have just described is self-limiting and thus explains the saturation of the 2DES density: each increase of V$_{\textrm{TG}}$ is directly compensated by an equivalent tunneling-assisted screening.\\
We have good experimental evidence for the fact, that the screening is only partial. To visualize this, one can consider Fig.~\ref{fig:figure5}(b) again, which already allowed to illustrate the loss of electrons in regimes II and III. We now consider V$_{\textrm{TG}}$ downsweeps, which started in regime IV (blue traces with starting points indicated by blue arrows). For any given V$_{\textrm{TG}}$ value in the range of $-1$ V and $+1$ V, the density always reacts immediately to a V$_{\textrm{TG}}$ downsweep, as shown by the three blue traces. While the density saturation illustrates the screening, the immediate reaction of the density demonstrates, that the screening is only partial.\\
While throughout the whole regime IV the system instantly reacts to gate action (although the density saturates), as V$_{\textrm{TG}}$ is increased further, a point occurs, from which the access to the 2DES, i.e. the capacitive coupling, is lost. This can be seen in Fig.~\ref{fig:figure5}(b), where when sweeping V$_{\textrm{TG}}$ between +1 V and +4 V (see the three green traces), either in an upsweep or a downsweep (downsweeps indicated by green arrows), the density remains constant. This suggests, that a regime V exists, here beyond V$_{\textrm{TG}} = +1$ V, in which the capacitive coupling to the 2DES is completely lost.
\paragraph*{Regime V}
In our understanding, this experimental observation translates that when a certain critical trapped electron density at the interface is reached with increasing V$_{\textrm{TG}}$, the partial screening (regime IV) transitions to full screening (regime V). Hence, the density in the 2DES cannot be adjusted anymore through variation of V$_{\textrm{TG}}$. This has also been reported in comparable gated heterostructures featuring thicker ($10$ nm) InGaAs cap layers \cite{Shabani.2014b}.\\
Note here, one peculiar observation in our experiments: while the charge density trapped at the interface is large enough to fully screen the action of V$_{\textrm{TG}}$, it does not contribute to the transport data. Even in this regime V, we do not detect any signature of parallel conduction. This is likely due to the previously mentioned hummocky potential landscape, that has formed at the interface. It seems plausible that such a potential landscape will provide puddle-like minima where charges will locally be trapped. These charges can, at best, hop between puddles and hence only negligibly contribute to the experimental transport signatures.
\paragraph*{Regime VI}
Conducting a TG downsweep after going through regimes I - V completes the largest observable gate hysteresis with a branch defined as regime VI, which we denoted as a linear regime with a smaller slope in Fig.~\ref{fig:figure1}, \ref{fig:figure2} and \ref{fig:figure3}. In Fig.~\ref{fig:figure3}, the slope of the branch in regime VI is significantly smaller than the slope of regime I. We observe that a settling time of the electron density on the scale of several tens of minutes after a variation of V$_{\textrm{TG}}$ is required in contrast to regime I, similar to the dynamics within regimes II, III and IV, indicating again a substantial charge transfer during regime VI. This motivates a closer look depicted in Fig.~\ref{fig:figure6}(b), where a TG downsweep at V$_{\textrm{TG}} = -1$ V was performed on sample B at V$_{\textrm{TG-BCd}} = -6$ V. Sweeping the TG down to V$_{\textrm{TG}} = -1.2$ V, the density is reduced, revealing the re-occurring of the capacitive coupling after regime V. Stopping at V$_{\textrm{TG}} = -1.2$ V and waiting for a given duration (horizontal line in the inset from Fig.~\ref{fig:figure6}(b)), however reveals a significant increase of the density over several minutes although V$_{\textrm{TG}}$ is kept constant (indicated by the vertical evolution of the density at V$_{\textrm{TG}} = -1.2$ V in Fig.~\ref{fig:figure6}(b)). Sweeping the TG up after this waiting time unveils a steeper slope for the linear density reaction compared to the slope of the previous downsweep. The density settles instantly in this upsweep and sweeping down leads to no hysteresis. This complex behavior in regime VI can be explained as follows: During the downsweep, a significant charge transfer counteracts the depletion caused by the TG, resulting in the smaller slope.
This charge transfer is a re-transfer of electrons, which were previously transferred from the QW towards the interface during the regimes II - V. We cannot distinguish during this re-transfer, whether participating electrons were trapped at the interface or at DDLs. Some of the trapped electrons are irreversibly trapped and cannot be brought back by a large negative V$_{\textrm{TG}}$ (see Fig.~\ref{fig:figure5}(c)), thus they do not contribute to this re-transfer in regime VI. As the QW is energetically favourable, these electrons accumulate in the QW, which we see in the measurement as a smaller slope and as a density increase during a waiting time (e.g. at V$_{\textrm{TG}} = -1.2$ V). 
Since the upsweep which follows a waiting time shows a steeper slope and a downsweep in this branch does not open up a hysteresis, the electron re-transfer towards the QW does not operate here. In fact, a hysteresis does not appear as long as a certain V$_{\textrm{TG}}$ (red circles in Fig.~\ref{fig:figure6}(b)) is not reached in a downsweep. This indicates, that the system is in an electrostatic metastable state. Fig.~\ref{fig:figure6}(c) shows an enlarged part of Fig.~\ref{fig:figure6}(b): When V$_{\textrm{TG}}$ reaches a red circle, the slope transitions from a larger (red dotted line in Fig.~\ref{fig:figure6}(c)) to a smaller (blue dotted line in Fig.~\ref{fig:figure6}(c)) slope value. This switching back to a smaller slope value hence marks the recommencement of the electron re-transfer into the QW. Notably, the position of the red circles (i.e. switching back to a smaller slope value) depends on the waiting time before sweeping. This can be seen in Fig.~\ref{fig:figure6}(c): The waiting time at V$_{\textrm{TG}} = -1.2$ V (vertical line in main graph, horizontal line in inset) is significantly larger than the one at V$_{\textrm{TG}} = -1.4$ V, resulting in a large $\Delta$V$_{\textrm{TG}}$ between the waiting time at V$_{\textrm{TG}} = -1.2$ V and the red circle (V$_{\textrm{TG}} \approx -1.35$ V) compared to the $\Delta$V$_{\textrm{TG}}$ left of the waiting time at V$_{\textrm{TG}} = -1.4$ V. Hence, the above mentioned electrostatic metastable state largely depends on the previous waiting time in this measurement sequence (downsweep, waiting time, upsweep, downsweep). As the nine sequences in Fig.~\ref{fig:figure6}(b) illustrate, the significant electron re-transfer into the QW is present throughout the whole regime VI until the MIT at V$_{\textrm{TG}} = -3.25$ V is reached. Hence, we show here that regime VI represents an unstable gating region where a depletion of the 2DES is constantly counteracted by electron re-transfer from the heterostructure into the 2DES.
\subsection{\label{ssec:temperature}Temperature dependence}
\begin{figure}[!htb]
\includegraphics{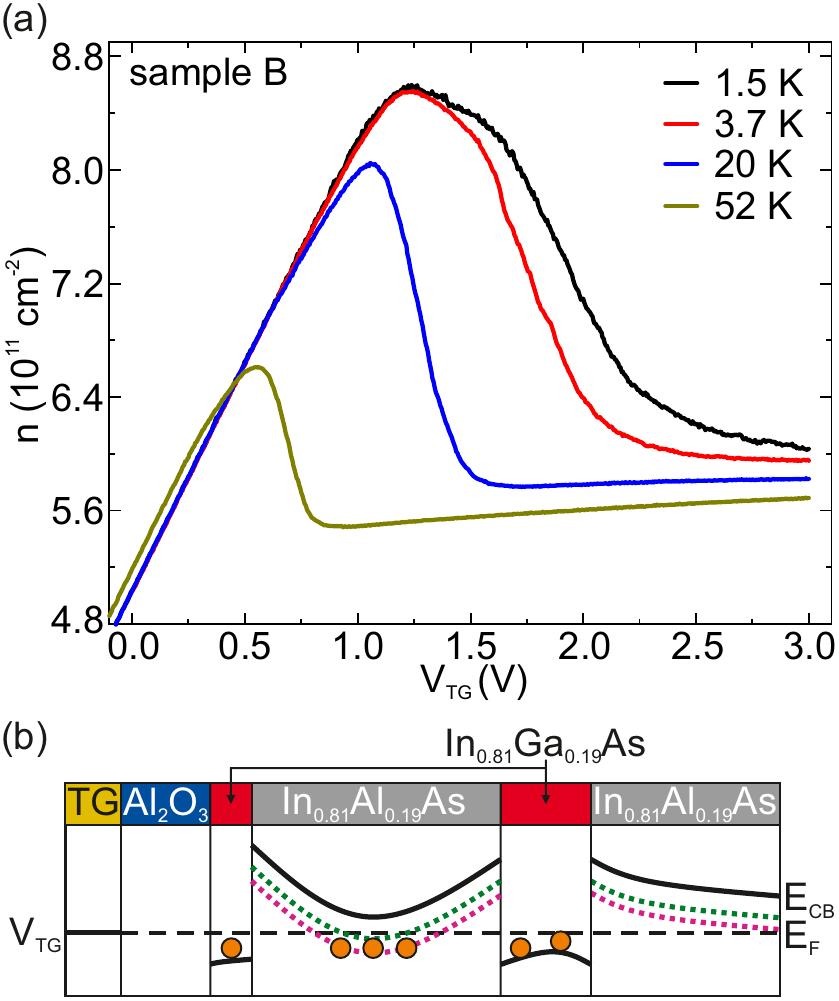}
\caption{\label{fig:figure7} (a) Electron density n of sample B as a function of V$_{\textrm{TG}}$ at V$_{\textrm{TG-BCd}} = 0$ V for four temperatures. With increasing thermal energy, the charge transfer processes are enhanced. (b) CB edge sketch of sample A (81\% indium content) at V$_{\textrm{TG}} = 0$ V, revealing the DDLs to lie below the Fermi energy. This results in a situation similar to regime IV already visible at V$_{\textrm{TG}} = 0$ V.}
\end{figure}

The charge transfer processes characteristic of regimes II, III and IV are based on electron tunnelling from the QW into the DDLs in the InAlAs. Hence, they should be enhanced by thermal assistance. Fig.~\ref{fig:figure7}(a) thus shows the electron density of sample B as a function of V$_{\textrm{TG}}$ at V$_{\textrm{TG-BCd}} = 0$ V for four different temperatures. The sample was initialized in regime I, a V$_{\textrm{TG}}$ sweep in positive direction then covers regimes I to IV. Note, that the sample was fully reset at room temperature between each measurement. The observed temperature dependence exactly matches the picture discussed in section C: the multi-step tunneling towards the interface via DDL sites, which marks the transition from regime III to IV, is enhanced by the slightest thermal assistance. This will result in a narrowing of the peak (transition from III to IV) as soon as the temperature is increased, as observed for all four temperatures in our experiment. Also, thermal assistance will render the triangular potential barrier for electrons sufficiently transparent to tunnel into the DDLs at smaller V$_{\textrm{TG}}$: the transition from regime I into regime II sets on for smaller V$_{\textrm{TG}}$. Hence, the peak height (onset of regime II) is reduced. This is indeed observed with increasing temperature (in particular blue and olive traces in Fig.~\ref{fig:figure7}(a)).
\subsection{\label{ssec:81indium}Discussion of the indium content}
Our experiments and the discussion in the previous sections point out the importance which the intrinsically present DDLs in InAlAs play in the observed limitations of the gated operation of InAlAs-based heterostructures. Hence, although this has not been addressed in the literature yet, we find that they represent a central parameter in the design of heterostructures for gated operation. Given the intrinsic nature of the DDLs, for example the indium concentration in the InAlAs as well as the depth of the QW below the heterostructure surface will strongly determine the limiting impact of the DDLs on the gate operation.\\
For example, the gate operation limitations of sample A (81\% indium) compared to sample B (75\% indium) reported in Fig.~\ref{fig:figure1} can consistently be explained within our model and be tracked down to the DDLs in InAlAs. The band edge schematic of sample A (81\% indium) is depicted in Fig.~\ref{fig:figure7}(b) and should be compared to Fig.~\ref{fig:figure4}(a) for sample B (75\% indium). We assume here, that the energy difference between the DDLs (dotted colored lines) and the InAlAs CB edge is independent of the indium concentration in InAlAs. However, the band offsets between InAlAs, InGaAs and InAs decrease with an increasing indium concentration. As a consequence, for high indium concentrations, tunneling from the QW into the DDLs will be enhanced already at V$_{\textrm{TG}} = 0$ V (see Fig.~\ref{fig:figure7}(b)), while it is forbidden for lower indium concentrations (see Fig.~\ref{fig:figure4}(a)). Two experimental features reported in our comparison of samples A and B in Fig.~\ref{fig:figure1} directly result from this enhanced tunneling: First, the maximal accessible electron density as well as the saturation is smaller in the sample with the higher indium concentration, sample A. Here, the onset of the multi-step hopping tunneling from the QW via InAlAs DDLs towards the interface (i.e. onsets of regimes II and IV) requires less gate action compared to a heterostructure with a lower indium concentration. Hence, when starting the gate operation in regime I (see Fig.~\ref{fig:figure1}), with similar electron densities, due to this earlier onset of multi-step tunneling, the maximum density (reached at the end of regime I) is smaller for sample A (81\% indium) than for sample B (75\% indium). The same is true for the saturation electron density observed in regimes IV and V (see Fig.~\ref{fig:figure1}). Second, for sample A, already at V$_{\textrm{TG}} = 0$ V, the reduced band offsets with increasing indium content induce an electron population in the area of the apex of the trough formed by the DDL energy levels in the schematic in Fig.~\ref{fig:figure7}(b), as indicated by the orange dots. As a consequence, when cooling down sample A at V$_{\textrm{TG}} = 0$ V, the gated system will be initialized in the unstable electrostatic regime VI, as observed in Fig.~\ref{fig:figure1}. As explained by our CTM, the device can only be brought into the stable regime I by applying sufficiently negative V$_{\textrm{TG}}$ to depopulate the DDLs. At the contrary, in sample B (lower indium concentration), negligible multi-step tunneling occurs at V$_{\textrm{TG}} = 0$ V (see Fig.~\ref{fig:figure4}(a)), even at RT. Hence, as discussed in Fig.~\ref{fig:figure1}, sample B is initialized in the stable gating regime I.\\
We have shown in Fig.~\ref{fig:figure2} that sample A may also be initialized in the stable regime I, provided that a negatively biased cooldown is applied (see green and orange traces in Fig.~\ref{fig:figure2}). This is in accordance with our CTM, since, by applying a sufficiently negative V$_{\textrm{TG-BCd}}$ at RT, the DDLs are depopulated as well as charges trapped at the interface. Then stable gate operation is allowed after the cooldown at this V$_{\textrm{TG-BCd}}$. Note that, as a direct consequence of this depopulation of charges, the biased cooldown brings the advantage of enabling a gate operation up to a higher electron density compared to the initialization at V$_{\textrm{TG}} = 0$ V, as seen from the comparison of the orange and green traces with the black one in Fig.~\ref{fig:figure2}. 
\section{\label{sec:conc}Conclusion}
In conclusion, our experimental study and charge transfer model identify the deep donor levels induced by defects intrinsically present in InAlAs to play a crucial role in the design process of InGaAs QWs embedded in InAlAs at indium contents $\geq75$\% e.g. for spin-orbitronic applications:
While providing native n-type doping to leverage the strong spin-orbit interaction in 2DES on the one hand, they at the same time substantially determine the electrostatics of the heterostructure through charge trapping and unintentional tunneling events. As a consequence, they thus impose limitations which need to be taken into account for a top-gated operation of 2DESs in such heterostructures. Strikingly, an electrostatically stable gate operation will quite generally only be possible in a limited gate voltage range. A much broader voltage range will then display reduced, to vanishing, capacitive coupling of the gate to the 2DES, induced by unstable or metastable electrostatic configurations, which are governed by charge transfer processes with long time scales.
We have shown that the design of the heterostructure - most importantly the indium content - crucially impacts in which electrostatic configuration a sample is initialized after cooldown. Our model allows to qualitatively predict how such heterostructures will react to gate operation and how to reach the electrostatically stable regime I. It also explains the observed favorable action of negative biased cooling to initialize a gated heterostructure in regime I with maximized accessible electron density. 
Hence, the presence of the deep donor levels should be accounted for in the design of heterostuctures for spin-orbitronics. In particular, trying to maximize the spin-orbit interaction through an increase of the indium content, will at the same time minimize the accessible electron density in gate operation.
\begin{acknowledgments}
We acknowledge the financial support of the Deutsche Forschungsgemeinschaft through Project ID 422 314695032-SFB1277 (Subproject A01). The authors thank D. Weiss for access to cleanroom facilities and technical assistance with temperature dependent measurements.
\end{acknowledgments}
\bibliography{bibliography.bib}
\end{document}